\documentstyle[epsfig,12pt,a4p]{article}

\newcommand{\pipipipi}{\mbox{$\pi^+\pi^-\pi^+\pi^-$ }}

\newcommand{\kkpi}{\mbox{$K^{0}_{S} K^{\pm} \pi^{\mp}$} }
\newcommand{\kkpip}{\mbox{$K^{0}_{S} K^{+} \pi^{-}$} }
\newcommand{\kkpim}{\mbox{$K^{0}_{S} K^{-} \pi^{+}$} }

\begin{document}
\begin{titlepage}
\pagestyle{empty}
\def\footnoterule{\hrule width 1.0\columnwidth}
\begin{tabbing}
put this on the right hand corner using tabbing so it looks
 and neat and in \= \kill
\> {9 July 1997}
\end{tabbing}
\bigskip
\bigskip
\begin{center}{\Large  {\bf A study of the $K \overline K \pi$ channel
produced centrally in pp interactions
at 450 GeV/c
}
}\end{center}
\bigskip
\bigskip
\begin{center}{        The WA102 Collaboration
}\end{center}\bigskip
\begin{center}{
D.\thinspace Barberis$^{  5}$,
W.\thinspace Beusch$^{   5}$,
F.G.\thinspace Binon$^{   7}$,
A.M.\thinspace Blick$^{   6}$,
F.E.\thinspace Close$^{  4}$,
K.M.\thinspace Danielsen$^{ 12}$,
A.V.\thinspace Dolgopolov$^{  6}$,
S.V.\thinspace Donskov$^{  6}$,
B.C.\thinspace Earl$^{  4}$,
D.\thinspace Evans$^{  4}$,
B.R.\thinspace French$^{  5}$,
T.\thinspace Hino$^{ 13}$,
S.\thinspace Inaba$^{   9}$,
A.V.\thinspace Inyakin$^{  6}$,
T.\thinspace Ishida$^{   9}$,
A.\thinspace Jacholkowski$^{   5}$,
T.\thinspace Jacobsen$^{  12}$,
G.V.\thinspace Khaustov$^{  6}$,
T.\thinspace Kinashi$^{  11}$,
J.B.\thinspace Kinson$^{   4}$,
A.\thinspace Kirk$^{   4}$,
W.\thinspace Klempt$^{  5}$,
V.\thinspace Kolosov$^{  6}$,
A.A.\thinspace Kondashov$^{  6}$,
A.A.\thinspace Lednev$^{  6}$,
V.\thinspace Lenti$^{  5}$,
S.\thinspace Maljukov$^{   8}$,
P.\thinspace Martinengo$^{   5}$,
I.\thinspace Minashvili$^{   8}$,
K.\thinspace Myklebost$^{   3}$,
T.\thinspace Nakagawa$^{  13}$,
K.L.\thinspace Norman$^{   4}$,
J.M.\thinspace Olsen$^{   3}$,
J.P.\thinspace Peigneux$^{  1}$,
S.A.\thinspace Polovnikov$^{  6}$,
V.A.\thinspace Polyakov$^{  6}$,
Yu.D.\thinspace Prokoshkin$^{\dag  6}$,
V.\thinspace Romanovsky$^{   8}$,
H.\thinspace Rotscheidt$^{   5}$,
V.\thinspace Rumyantsev$^{   8}$,
N.\thinspace Russakovich$^{   8}$,
V.D.\thinspace Samoylenko$^{  6}$,
A.\thinspace Semenov$^{   8}$,
M.\thinspace Sen\'{e}$^{   5}$,
R.\thinspace Sen\'{e}$^{   5}$,
P.M.\thinspace Shagin$^{  6}$,
H.\thinspace Shimizu$^{ 14}$,
A.V.\thinspace Singovsky$^{  6}$,
A.\thinspace Sobol$^{   6}$,
A.\thinspace Solovjev$^{   8}$,
M.\thinspace Stassinaki$^{   2}$,
J.P.\thinspace Stroot$^{  7}$,
V.P.\thinspace Sugonyaev$^{  6}$,
K.\thinspace Takamatsu$^{ 10}$,
G.\thinspace Tchlatchidze$^{   8}$,
T.\thinspace Tsuru$^{   9}$,
G.\thinspace Vassiliadis$^{\dag   2}$,
M.\thinspace Venables$^{  4}$,
O.\thinspace Villalobos Baillie$^{   4}$,
M.F.\thinspace Votruba$^{   4}$,
Y.\thinspace Yasu$^{   9}$.
}\end{center}

\begin{center}{\bf {{\bf Abstract}}}\end{center}

{
Results are presented of an analysis of the
reactions
$ pp \rightarrow p_f$(\kkpi)$p_s$
and \break
$ pp \rightarrow p_f(K^0_SK^0_S \pi^0)p_s$
at 450 GeV/c.
Clear $f_1$(1285) and $f_1$(1420) signals are seen and
a spin parity analysis
shows that both have $I^GJ^{PC}$~=~$0^+1^{++}$.
The $f_1$(1285) decays to $a_0(980) \pi$ and
the $f_1$(1420) decays to $K^* \overline K $.
Both states have a similar dependence as a function of $dP_T$ consistent with
what has been observed for other $q \overline q$ states.
Evidence is also presented for a $K^* \overline K$ decay mode of the
$\eta_2(1620)$.
}
\bigskip
\bigskip\begin{center}{{Submitted to Physics Letters}}
\end{center}
\bigskip
\bigskip
\begin{tabbing}
aba \=   \kill
$^\dag$ \> \small
Deceased. \\
$^1$ \> \small
LAPP-IN2P3, Annecy, France. \\
$^2$ \> \small
Athens University, Nuclear Physics Department, Athens, Greece. \\
$^3$ \> \small
Bergen University, Bergen, Norway. \\
$^4$ \> \small
School of Physics and Astronomy, University of Birmingham, Birmingham, U.K. \\
$^5$ \> \small
CERN - European Organization for Nuclear Research, Geneva, Switzerland. \\
$^6$ \> \small
IHEP, Protvino, Russia. \\
$^7$ \> \small
IISN, Belgium. \\
$^8$ \> \small
JINR, Dubna, Russia. \\
$^9$ \> \small
High Energy Accelerator Research Organization (KEK), Tsukuba, Ibaraki 305,
Japan. \\
$^{10}$ \> \small
Faculty of Engineering, Miyazaki University, Miyazaki, Japan. \\
$^{11}$ \> \small
RCNP, Osaka University, Osaka, Japan. \\
$^{12}$ \> \small
Oslo University, Oslo, Norway. \\
$^{13}$ \> \small
Faculty of Science, Tohoku University, Aoba-ku, Sendai 980, Japan. \\
$^{14}$ \> \small
Faculty of Science, Yamagata University, Yamagata 990, Japan. \\
\end{tabbing}
\end{titlepage}
\newpage
\pagestyle{plain}
\setcounter{page}{2}
In previous analyses~\cite{OLDE1,OLDE2,FURTHERE}
of the centrally produced \kkpi system
the peaks observed at 1.28 and 1.42 GeV were found to have
$J^{PG}$~=~$1^{++}$ and hence were identified with the
$f_1$(1285) and $f_1(1420)$ respectively.
However, due to the limited statistics
a 10$\%$ $0^{-+}$ contribution could not be excluded~\cite{FURTHERE}.
\par
Originally
the $f_1(1420)$ was thought to be the $s \overline s$ isoscalar member of
the ground state $1^{++}$ nonet, the other members being
the $a_1(1260)$ triplet, the $K_1(1270/1400)$ and the $f_1(1285)$.
The $f_1(1420)$ was found to decay dominantly to $K^* \overline K$ hence
reinforcing its $s \overline s $ assignment.
It is commonly accepted that
$ s \overline s$ objects should be preferentially produced
in $K^-$ incident experiments whereas
in the study of the reaction
\begin{center}
$ K^- p$ $ \rightarrow$ \kkpi $\Lambda$
\end{center}
two experiments
\cite{GAV,LASS}
observed only weak evidence for a $f_1$(1420) signal. Instead
they found evidence for a new
J$^{PC}$~=~$1^{++}$ state with a mass of 1.53 GeV and a width of 100 MeV,
called the D$^\prime$/$f_1(1510)$.
It was suggested that this state is a better
candidate for the $ s \overline s$  member of the $1^{++}$ nonet based on
its production.
\par
Therefore, the $1^{++}$ nonet appears to have ten
candidates with the $f_1$(1420)
thought to be the extra state.
As a $1^{++}$ state
its mass is considered to be too low to be a glueball or hybrid state,
and it has been suggested that it
could be either a hybrid~\cite{ISHIDA}, a four quark state
\cite{FOURQUARK}
or a $K^* \overline K$ molecule
\cite{WEINSTEIN,LONGACRE}.
\par
This paper presents a study of
the centrally produced exclusive final states formed in the reactions
\begin{equation}
pp \rightarrow p_{f} (K^0_S K^{\pm}\pi^{\mp}) p_{s}
\label{eq:a}
\end{equation}
and
\begin{equation}
pp \rightarrow p_{f} (K^0_S K^0_S\pi^0) p_{s}
\label{eq:b}
\end{equation}
at 450 GeV/c,
where the subscripts $f$ and $s$ indicate the fastest and slowest particles
in the laboratory respectively.
The data presented here represent a factor of ten increase compared
to previously published data.
In addition, in order to try to gain more information about the nature
of the $f_1(1420)$,
a study is performed
as a function of $dP_T$. This variable is the difference
in the transverse momentum vectors of the two exchanged particles~\cite{WADPT}
and has been proposed as a glueball-$q \overline q$ filter~\cite{ck97}.
\par
The data come from the 1995 and 1996 runs of experiment WA102
which has been performed using the CERN Omega Spectrometer.
The layout of the Omega Spectrometer used in these runs is similar to that
described in ref.~\cite{wa9192} with the replacement of the
OLGA calorimeter by GAMS~4000~\cite{gams}.
In the 1996 run of the experiment the setup was augmented by the addition
of a threshold {\v C}erenkov counter for charged particle identification.
\par
Reaction (1)
has been isolated from the sample of events having four outgoing
tracks plus a reconstructed $V^0$
by first imposing on the components of missing momentum
the cuts $| \Delta P_x |$~$\leq$~14.0~GeV/c,
$| \Delta P_y |$~$\leq$~0.12~GeV/c and
$| \Delta P_z |$~$\leq$~0.08~GeV/c,
where the x axis is along the beam direction.
The $\pi^+ \pi^-$  mass distribution for the $V^0$s shows a clear $K^0$
signal which
was selected by requiring 0.475 $\leq$ $m(\pi^+ \pi^-)$ $\leq$ 0.520~GeV.
Reaction (1)
was then selected from this sample
by using energy conservation. A cut of $|\Delta|$~$\leq$~1.6~(GeV)$^2$
was used, where
$\Delta = MM^2(p_f p_s) - M^2(K^0_S K^{\pm} \pi^{\mp})$.
\par
These cuts gave
an ambiguous
\kkpi assignment
to 59$\%$ of the events.
The Ehrlich mass
\cite{EHRLICH}
has been calculated for the $V^0$ and one of the charged
particles, assuming the other
to be a pion. A clear peak is observed in the Ehrlich mass spectrum
(not shown)
at the kaon mass squared and
suitable cuts were applied
to select out the
\kkpip and \kkpim channels.
\par
Fig.~\ref{kkpi}a) shows the \kkpi effective mass spectrum
(52 141 events)
where the events that still
have an ambiguous mass assignment
are plotted twice (18~$\%$~of the events).
Fig.~\ref{kkpik}a) shows the \kkpi effective mass spectrum where the kaon
has been positively identified by the {\v C}erenkov counter
(5 647 events).
A fit to these spectra,
using two Breit-Wigners convoluted with Gaussians to
account for the experimental resolution ($\sigma$~=~10~MeV
in the $f_1(1285)$ region and
$\sigma$~=~13~MeV
in the $f_1(1420)$ region)
and a background
of the form $a(m-m_{th})^b exp(-cm-dm^2)$, where
$m$ is the \kkpi mass,
$m_{th}$ is the threshold mass and
a,b,c,d are fit parameters,
yields masses and widths of
\begin{tabbing}
abaabddb \= mdsd \= mm \=1234 pm 23mm\= mmmyyyy  \= gam \= mm \= 312 pm12 \=gev
 \kill
\> $M_1$ \>= \>1281 $\pm$  1 \> MeV \>$\Gamma_1$  \>= \>20 $\pm$ 2 \>MeV \\
\> $M_2$ \>= \>1426 $\pm$ 1 \> MeV \>$\Gamma_2$  \>= \>58 $\pm$ 4 \>MeV.
\end{tabbing}
\par
A Dalitz plot analysis
of the \kkpi final state has been performed
using Zemach tensors and a standard isobar model
\cite{oldpap}.
The \kkpi mass spectrum has been fitted in 20~MeV slices from
1.23 to 1.93~GeV.
Different combinations of 22 waves,
with J~$\leq$~2, decaying to $a_0(980) \pi$ and $K^* \overline K$ have been
used,
where the
$a_0(980)$ has been described in terms of the Flatt\'{e} formalism
\cite{flattee}.
Full interference between waves having the same spin-parity
has been allowed. Interference was also allowed between the
1$^{++}$ and 1$^{+-}$ $K^* \overline K$ waves.
\par
The analysis has first been performed on the data without particle
identification. This has the advantage of having the greatest statistics
but has the disadvantage that 18~$\%$ of the events have a
$K^{\pm}\pi^{\mp}$ ambiguity.
This ambiguity does not affect the analysis of the 1.28 and 1.42~GeV regions
but means that higher masses are contaminated from wrong sign reflections
of the dominant peaks.
The fit shows that
the
$J^{PG}$~=~1$^{++}$ $a_0(980) \pi$ wave
is the only wave required to describe the $f_1$(1285) peak and
that
the $J^{PG}$~=~1$^{++}$ $K^* \overline K$ wave
is the only wave required to describe the $f_1$(1420) peak.
The addition of 0$^{-+}$ or any other
waves up to J equals 2 does not increase the
likelihood significantly.
The results of the fit
are shown in
fig.\ref{kkpi} b),c) and d).
It is interesting to note that there is no evidence for the
$f_1(1510)$.
\par
A spin analysis has also been performed on the sample where the $K$ is
identified by the {\v C}erenkov system.
This sample is lower in statistics but has the advantage that
there are no ambiguities and hence the background at higher masses is reduced.
The result of the analysis in the 1.28 and 1.42~GeV regions is the same as
for the sample without kaon identification;
only the $J^{PG}$~=~$1^{++}$
waves are required in the fit. In addition, a
$J^{PG}$~=~$2^{-+}$ $K^* \overline K$ wave
is found to be required in the 1.6~GeV region as shown in fig~\ref{kkpik}e).
The addition of this wave increases the log likelihood by
$\Delta{\cal L} = 15$  corresponding to
$n=\sqrt{2\Delta{\cal L}}=5.5$ standard deviations.
Superimposed on
fig~\ref{kkpik}e) is the Breit-Wigner found to describe the $\eta_2(1620)$
decaying to $a_2(1320) \pi$ in the analysis of the \pipipipi
final state of this experiment~\cite{wa1024pi}. As can be seen the
$J^{PG}$~=~$2^{-+}$ $K^* \overline K$ wave is consistent with this in mass
and width.
\par
{}From a Dalitz plot analysis of the \kkpi system only the $J^{PG}$
of the final state can be determined. In order to determine the C parity,
and hence the isospin of the final states, a study of the
$K^0_S K^0_S \pi^0$ system is required since only C parity positive
states can be observed in this channel.
\par
Reaction (2)
has been isolated from the sample of events having two outgoing
tracks together with two reconstructed $V^0$s and a $\pi^0$
decaying to 2$\gamma$s which were detected in the
GAMS calorimeter
by first imposing the cuts on the components of missing momentum
 described above.
The $\pi^+ \pi^-$  mass distributions for the $V^0$s show clear $K^0$
signals which
were selected by requiring 0.475 $\leq$ $m(\pi^+ \pi^-)$ $\leq$ 0.520~GeV.
\par
Fig.~\ref{k0k0pi0}a) shows the $K^0_S K^0_S \pi^0$ effective mass
spectrum (949 events) where clear peaks at 1.28 and 1.42~GeV can be seen.
A spin analysis has been performed on this sample
in 40~MeV slices from 1.21 to 1.69~GeV
and shows that the peaks at 1.28 and 1.42 GeV have $J^{PC}$~=~$1^{++}$.
The small number of events does not allow a statistically significant
signal in the
$J^{PC}$~=~$2^{-+}$ wave to be extracted.
\par
Therefore, we have clearly identified both the $f_1(1285)$ and the
$f_1(1420)$ to have
$I^GJ^{PC}$ =~$0^+1^{++}$. The $f_1(1285)$ is found to
decay only to $a_0(980)\pi$ while the $f_1(1420)$ is found to
decay only to $K^* \overline K$. There is also evidence for a
$J^{PG}$~=~$2^{-+}$ $K^* \overline K$ signal consistent with the
$\eta_2(1620)$.
\par
The $f_1(1285)$ has also been observed in the $\pi^+\pi^-\pi^+\pi^-$
channel of this experiment~\cite{wa1024pi} and its
branching ratio to $K \overline K \pi$ and
$4\pi$ has been calculated taking into account the unseen decay modes
and gives
\begin{equation}
\frac{f_1(1285) \rightarrow K \overline K \pi}{f_1(1285) \rightarrow 4\pi} =
0.265 \pm 0.01 \pm 0.01.
\end{equation}
This is a considerable improvement in accuracy on our previous
determination~\cite{old4pi}.
\par
Assuming the signal in the
$J^{PC}$~=~$2^{-+}$ $K^* \overline K$ wave to be due to the $\eta_2(1620)$
observed decaying to $a_2(1320)\pi$ in the $\pi^+\pi^-\pi^+\pi^-$
channel~\cite{wa1024pi}
the branching ratio for the $\eta_2(1620)$ to $K \overline K \pi$ and
$a_2(1320) \pi$ has been calculated taking into account the unseen decay modes
and gives
\begin{equation}
\frac{\eta_2(1620) \rightarrow K \overline K \pi}{\eta_2(1620) \rightarrow
a_2(1320) \pi} = 0.07  \pm 0.02 \pm 0.02.
\end{equation}
\par
Since no signal is observed for the $f_1(1510)$ in the
$1^{++}$ $K^* \overline K$ wave, an upper limit for its
production in central collisions relative to the $f_1(1420)$ has been
calculated and gives
\begin{equation}
\frac{f_1(1510) \rightarrow K^* \overline K}{f_1(1420) \rightarrow K^*
\overline K}  < 0.03 \;\;\;\;\;\;\;\;\;\;\;  at\:90 \% \: CL.
\end{equation}
\par
Close and Kirk~\cite{ck97}
have proposed that when the centrally produced system is analysed
as a function of the parameter $dP_T$, which is the difference
in the transverse momentum vectors of the two exchange particles~\cite{WADPT},
states with large (small) internal angular momentum will be enhanced at
large (small) $dP_T$.
A study of the \kkpi mass spectrum as a function of $dP_T$ is presented in
fig.~\ref{kkpidpt}a), b) and c)
for $dP_T$~$\leq$~0.2~GeV,
0.2~$\leq$~$dP_T$~$\leq$~0.5~GeV and
$dP_T$~$\geq$~0.5~GeV respectively.
As can be seen both the
$f_1$(1285) and $f_1(1420)$ signals behave similarly, which
is consistent with both states having the same underlying dynamical
structure.
Table~\ref{ta:1} gives the percentage of each resonance in the three
$dP_T$ regions considered
and shows a strong suppression of both resonances at
small $dP_T$ similar to that found for other $q \overline q$
states~\cite{WADPT}.
\par
In conclusion,
clear $f_1$(1285) and $f_1$(1420) signals are observed.
A spin parity analysis
shows that both are $I^GJ^{PC}$~=~$0^+1^{++}$
states.
The $f_1$(1285) is found to decay via $a_0(980)\pi$ while
the $f_1$(1420) is found to decay only to $K^* \overline K$;
no 0$^{-+}$ or 1$^{+-}$ waves are required to describe the data.
The $dP_T$ dependence of both states is similar and is consistent
with both states having a $q \overline q$ nature.
There is also evidence for a $K^* \overline K$ decay mode of the
$\eta_2(1620)$.

\newpage
{ \large \bf Tables \rm}
\begin{table}[h]
\caption{Resonance production as a function of $dP_T$
expressed as a percentage of its total contribution.}
\label{ta:1}
\vspace{1in}
\begin{center}
\begin{tabular}{|c|c|c|c|} \hline
 & & &  \\
 &$dP_T$$\leq$0.2 GeV & 0.2$\leq$$dP_T$$\leq$0.5 GeV &$dP_T$$\geq$0.5 GeV\\
 & & & \\ \hline
 & & & \\
$f_{1}(1285)$  &7.2 $\pm$ 1.2 & 53.4 $\pm$ 2.7 &39.4 $\pm$ 2.0 \\
 & & & \\ \hline
 & & & \\
$f_{1}(1420)$  &4.5 $\pm$ 1.2 & 51.3 $\pm$ 2.0 &44.2 $\pm$ 1.5 \\
 & & & \\ \hline
\end{tabular}
\end{center}
\end{table}
\newpage
{ \large \bf Figures \rm}
\begin{figure}[h]
\caption{The full \kkpi sample. a) Mass spectrum with fit described in the
text,
b) the \kkpi mass spectrum used in the spin analysis and the resulting
phase space contribution,
c) the $J^{PG}=1^{++}$ $a_0(980)\pi$ wave and
d) the $J^{PG}=1^{++}$ $K^* \overline K$ wave.
}
\label{kkpi}
\end{figure}
\begin{figure}[h]
\caption{The \kkpi data with the $K^{\pm}$ identified by the
{\v C}erenkov counter. a) Mass spectrum with fit described in the text,
b) the \kkpi mass spectrum used in the spin analysis and the resulting
phase space contribution,
c) the $J^{PG}=1^{++}$ $a_0(980)\pi$ wave,
d) the $J^{PG}=1^{++}$ $K^* \overline K$ wave and
e) the $J^{PG}=2^{-+}$ $K^* \overline K$ wave.
}
\label{kkpik}
\end{figure}
\begin{figure}[h]
\caption{The $K^0_SK^0_S\pi^0$ sample. a) Mass spectrum,
b) the $K^0_SK^0_S\pi^0$ mass spectrum
used in the spin analysis and the resulting
phase space contribution,
c) the $J^{PC}=1^{++}$ $a_0(980)\pi$ wave and
d) the $J^{PC}=1^{++}$ $K^* \overline K$ wave.
}
\label{k0k0pi0}
\end{figure}
\begin{figure}[h]
\caption{The \kkpi mass spectrum as a function of $dP_T$ for
a) $dP_T$~$\leq$~0.2,
b) 0.2~$\leq$~$dP_T$~$\leq$~0.5 and
c) $dP_T$~$\geq$~0.5~GeV.
}
\label{kkpidpt}
\end{figure}
\newpage
\begin{center}
\epsfig{figure=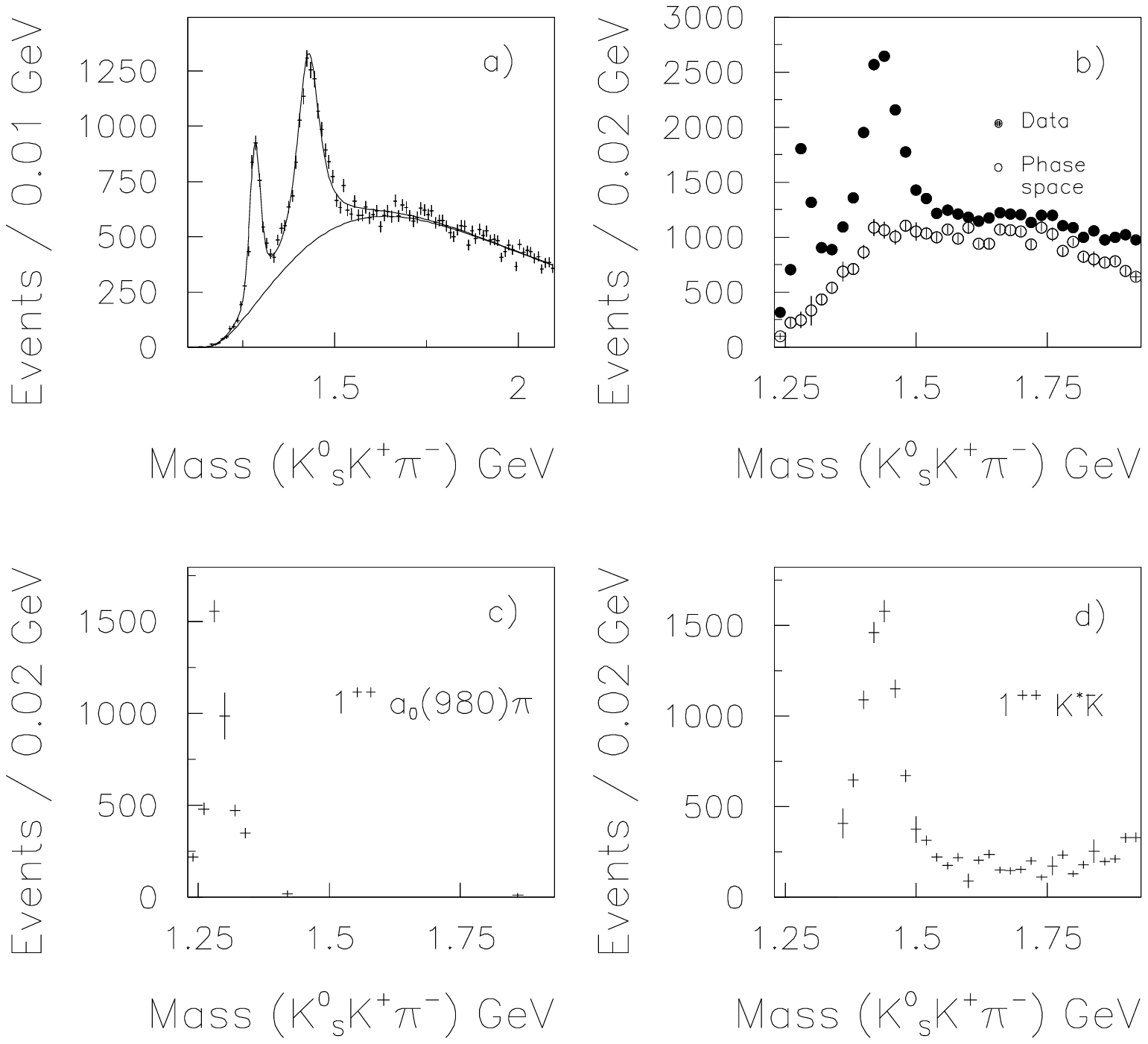,height=22cm,width=17cm}
\end{center}
\begin{center} {Figure 1} \end{center}
\newpage
\begin{center}
\epsfig{figure=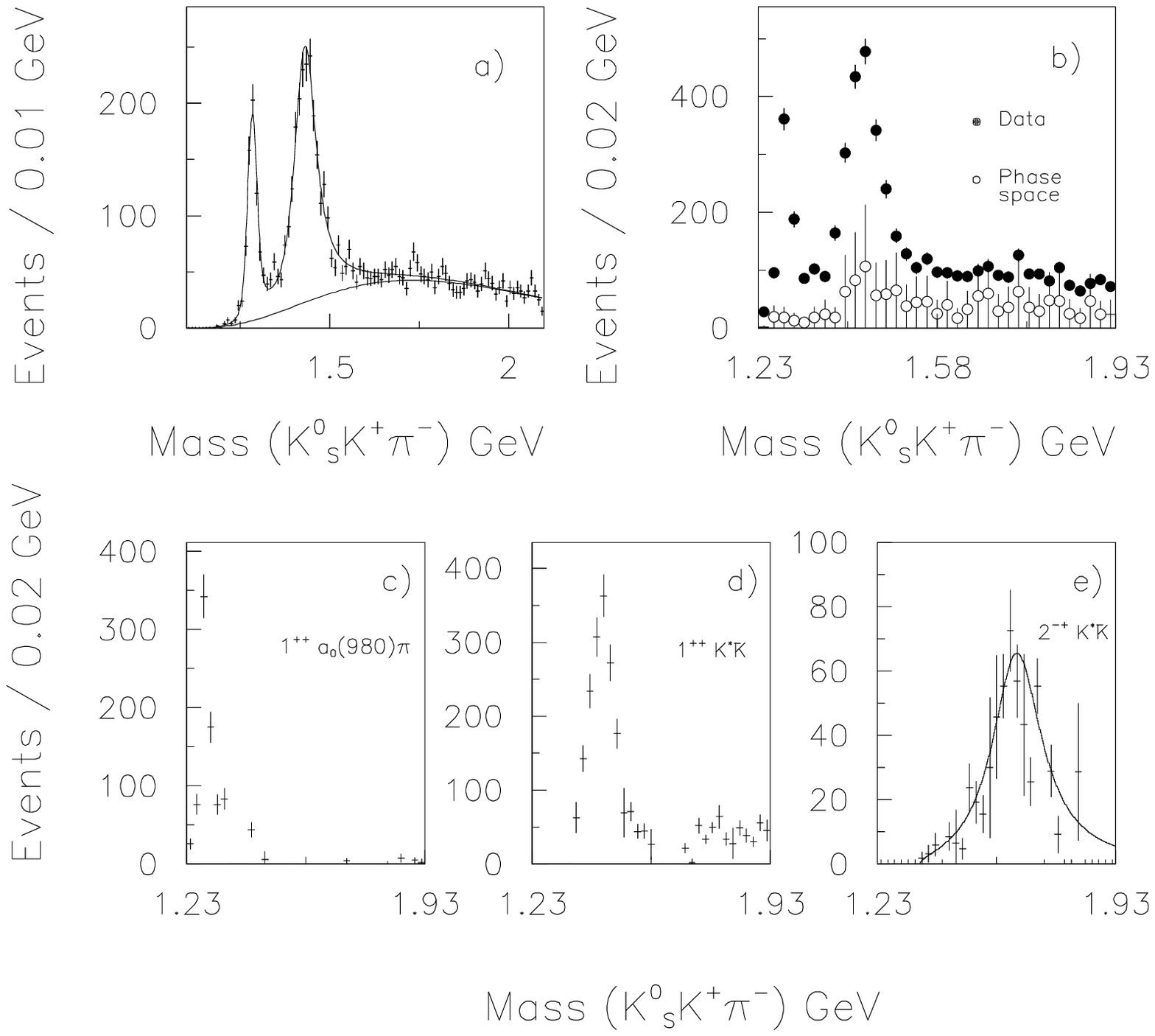,height=22cm,width=17cm}
\end{center}
\begin{center} {Figure 2} \end{center}
\newpage
\begin{center}
\epsfig{figure=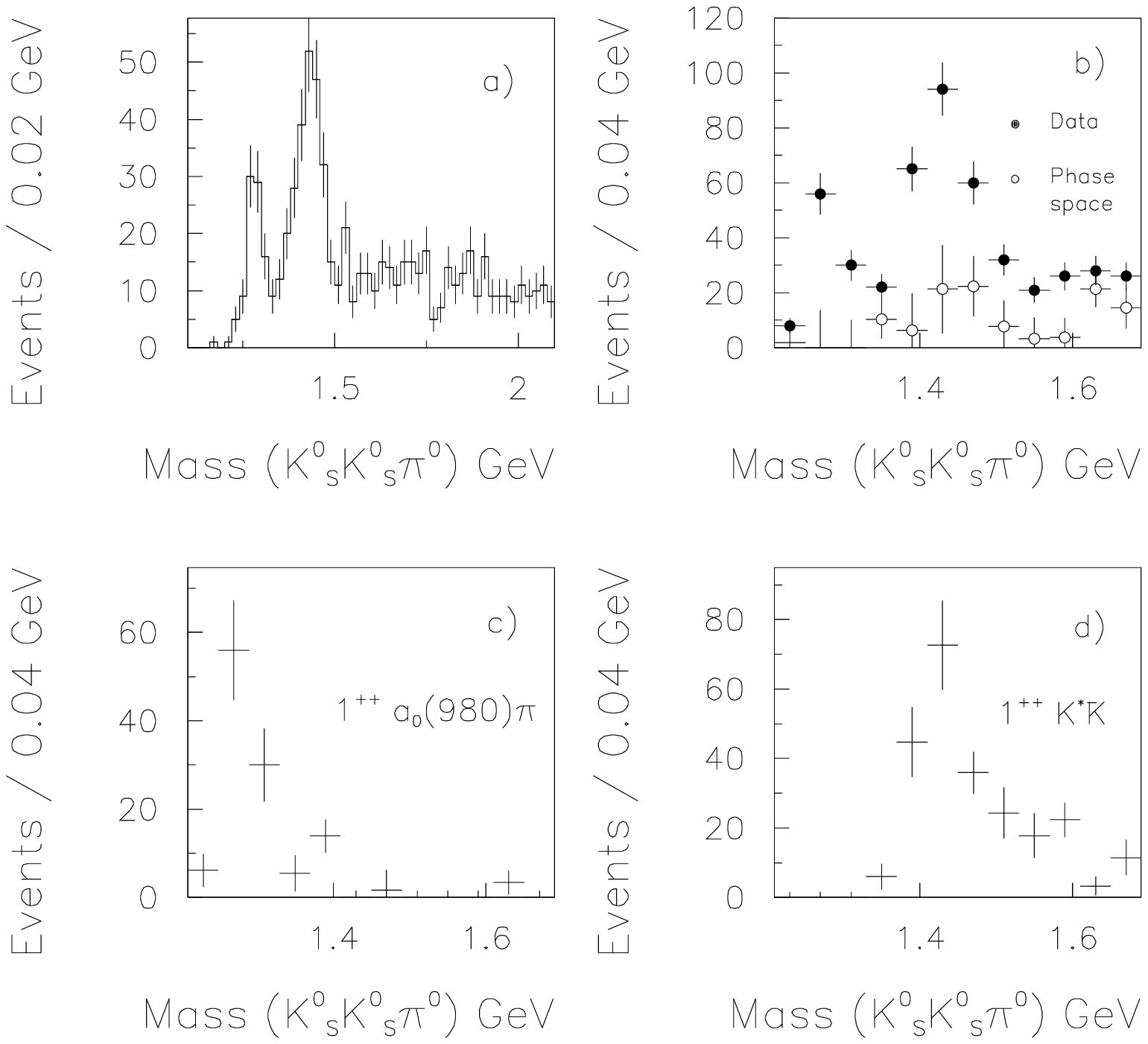,height=22cm,width=17cm}
\end{center}
\begin{center} {Figure 3} \end{center}
\newpage
\begin{center}
\epsfig{figure=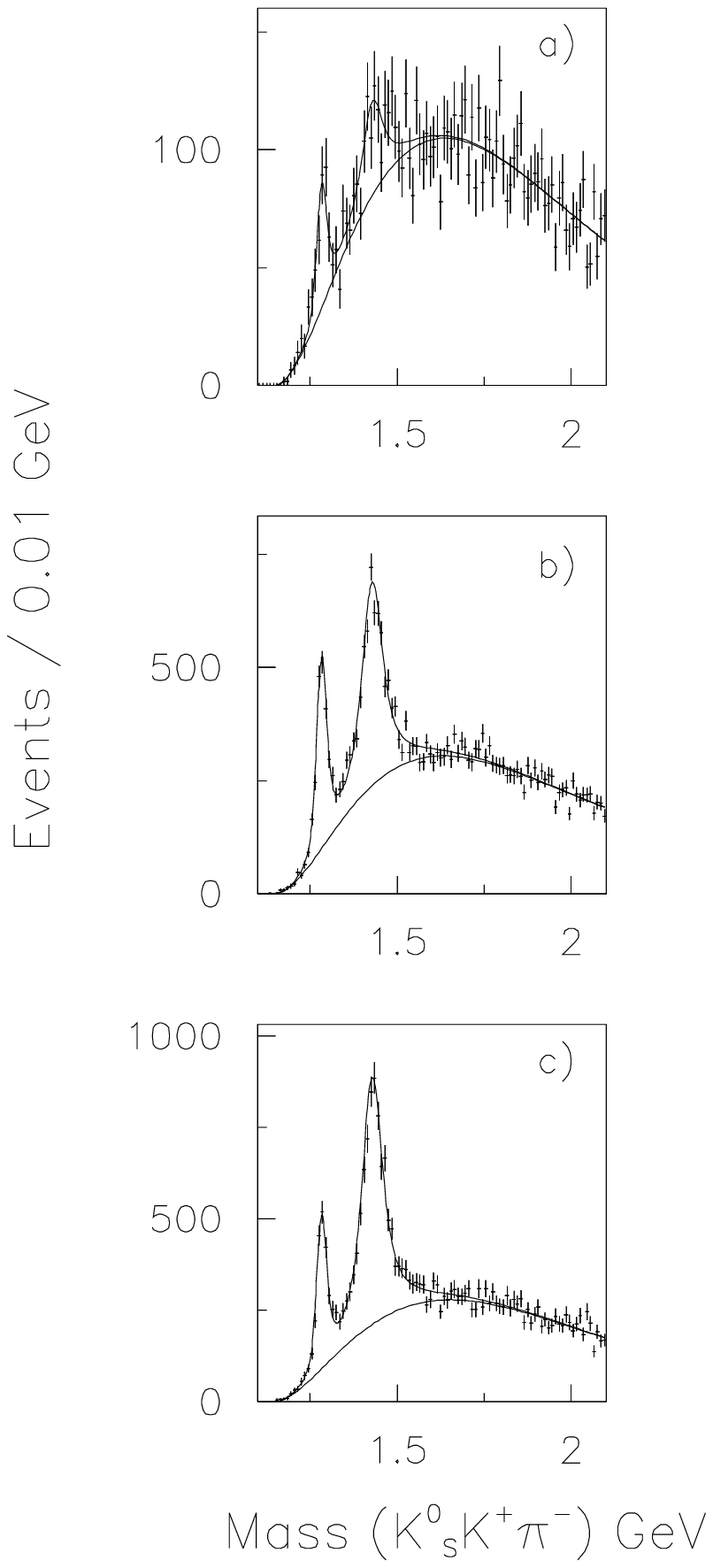,height=22cm,width=17cm}
\end{center}
\begin{center} {Figure 4} \end{center}
\end{document}